\documentclass[aps,twocolumn,preprintnumbers,showpacs,showkeys]{revtex4}
\usepackage{epsfig}
\usepackage{amssymb,amsmath,amsfonts,amsthm,graphicx}
\graphicspath{{./Figures/}}                 
\def\Pol{{\mathcal{P}}}
\def\diag{{\rm diag}}

\newcommand{\beq}{\begin{equation}}
\newcommand{\eeq}{\end{equation}}
\newcommand{\bea}{\begin{array}}
\newcommand{\eea}{\end{array}}

\newcommand{\Sec}[1]{Section~\ref{#1}}
\newcommand{\Eq}[1]{Eq.~(\ref{#1})}

\newcommand{\tr}{\operatorname{Tr}}

\newcommand{\half}{\frac{1}{2}}
\newcommand{\veps}{\varepsilon}

\def\cA{{\cal{A}}}
\def\cT{{\hat{\beta}}}
\def\cT{{\beta}}

\def\beqa{\begin{eqnarray}}
\def\eeqa{\end{eqnarray}}
\def\pl{{{\cal P}_\infty}}
\def\plo{{{\cal P}_\infty^0}}
\def\tr{{\rm tr}}
\def\Tr{{\rm Tr}}
\def\cD{{\cal{D}}}
\def\cO{{\cal{O}}}
\newcommand{\basispl}{
   \put(-.5,-.5){\line(1,0){1}}
   \put(.5,-.5){\line(0,1){1}}
   \put(.5,.5){\line(-1,0){1}}
   \put(-.5,.5){\line(0,-1){1}}
                         }
\newcommand{\basisar}{
   \put(0,-.5){\vector(1,0){0}}
   \put(.5,0){\vector(0,1){0}}
   \put(0,.5){\vector(-1,0){0}}
   \put(-.5,0){\vector(0,-1){0}}
                  }
\newcommand{\plaq}{\setlength{\unitlength}{.5cm}\raisebox{-.2cm}{
   \begin{picture}(1.2,1.2)(-.6,-.6)
   \basispl\basisar
   \put(-.5,-.5){\circle*{.2}}
   \put(-.55,-.55){\makebox(0,0)[tr]{\footnotesize $x$}}
   \put(-.55,0){\makebox(0,0)[r]{\footnotesize $\nu$}}
   \put(0,-.55){\makebox(0,0)[t]{\footnotesize $\mu$}}
   \end{picture}}}

\newcommand{\twoplaq}{\setlength{\unitlength}{1cm}\raisebox{-.5cm}{
   \begin{picture}(1.2,1.2)(-.6,-.6)
   \basispl
   \put(-.5,-.5){\circle*{.1}}
   \put(-.5,.5){\circle*{.1}}
   \put(.5,-.5){\circle*{.1}}
   \put(.5,.5){\circle*{.1}}
   \put(0,-.5){\circle*{.1}}
   \put(0,.5){\circle*{.1}}
   \put(.5,0){\circle*{.1}}
   \put(-.5,0){\circle*{.1}}
   \put(-.25,-.5){\vector(1,0){0}}
   \put(.25,-.5){\vector(1,0){0}}
   \put(.5,-.25){\vector(0,1){0}}
   \put(.5,.25){\vector(0,1){0}}
   \put(-.25,.5){\vector(-1,0){0}}
   \put(.25,.5){\vector(-1,0){0}}
   \put(-.5,-.25){\vector(0,-1){0}}
   \put(-.5,.25){\vector(0,-1){0}}
   \put(-.55,-.55){\makebox(0,0)[tr]{\footnotesize $x$}}
   \put(-.55,0){\makebox(0,0)[r]{\footnotesize $\nu$}}
   \put(0,-.55){\makebox(0,0)[t]{\footnotesize $\mu$}}
   \end{picture}}}

\newcommand{\linkhmu}{\setlength{\unitlength}{.5cm}\raisebox{-.2cm}{
   \begin{picture}(1.2,1.2)(-.6,-.6)
   \put(.5,0){\line(-1,0){1}}
   \put(0,0){\vector(1,0){0.1}}
   \put(-.5,0){\circle*{.2}}
   \put(-.35,-.25){\makebox(0,0)[tr]{\footnotesize $x$}}
   \put(0.4,-.3){\makebox(0,0)[r]{\footnotesize $\mu$}}
   \end{picture}}}

\begin{document}

\preprint{ITEP-LAT/2013-02,~HU-EP/13-09}

\title{Topology across the finite temperature transition
studied by overimproved cooling in gluodynamics and QCD }

\author{V.~G.~Bornyakov}
\affiliation{Institute for High Energy Physics, 142 281 Protvino, Russia \\
and Institute of Theoretical and Experimental Physics, 117259 Moscow, Russia}

\author{E.-M. Ilgenfritz}
\affiliation{Joint Institute for Nuclear Research, VBLHEP, 141980 Dubna, Russia}

\author{B.~V.~Martemyanov}
\affiliation{
Institute of Theoretical and Experimental Physics, 117259 Moscow, Russia\\
National Research Nuclear University MEPhI, 115409, Moscow, Russia\\
Moscow Institute of Physics and Technology, 141700, Dolgoprudny, Moscow Region,
Russia}

\author{V.~K.~Mitrjushkin}
\affiliation{Joint Institute for Nuclear Research, BLTP, 141980 Dubna, Russia \\
and Institute of Theoretical and Experimental Physics, 117259 Moscow, Russia}

\author{M.~M\"uller-Preussker}
\affiliation{Humboldt-Universit\"at zu Berlin, Institut f\"ur Physik,
  12489 Berlin, Germany}
\date{April 3, 2013}

\begin{abstract}
Gluodynamics and two-flavor QCD at non-zero temperature are studied
with the so-called overimproved cooling technique under which caloron 
solutions may remain stable.
We consider topological configurations either at the first occuring stable
plateau of topological charge or at the first (anti)selfdual plateau and 
find the corresponding topological susceptibility at various temperatures 
on both sides of the thermal transition or crossover. In pure gluodynamics 
the topological susceptibility drops sharply at the deconfinement temperature 
while in full QCD it decreases smoothly at temperatures above the 
pseudocritical one. The results are close to those calculated by other methods.
We interpret our findings in terms of the (in)stability of calorons with 
non-trivial holonomy and their dyon constituents against overimproved cooling.
\end{abstract}



\maketitle

\section{Introduction}
\label{sec:introduction}

Topological charge has many faces. It is well known that the appearance of its 
space-time distribution strongly depends on the method by which it is studied.
If resolved by means of the modes of a lattice Dirac operator with maximal
chiral symmetry (the overlap Dirac operator), the resolution can be dialled 
by the cutoff applied to the eigenvalues of the modes (in a symmetric band 
enclosing the zero eigenvalue) included in that analysis. 
In this way very different types of topological structures are revealed, ranging 
from globally extended, laminar sheets of alternating sign to instanton-like lumps 
appearing in the infrared (IR)~\cite{Horvath:2001ir, Horvath:2003yj,
Ilgenfritz:2007xu,Ilgenfritz:2008ia,Thacker:2012nx}.
To some extent the scale of ultraviolet
filtering can be mimicked by the number, say, of overimproved stout link smearing 
steps~\cite{Ilgenfritz:2008ia}. Before the fermionic methods became popular, only 
cooling~\cite{Ilgenfritz:1985dz} and smearing~\cite{Albanese:1987ds} were available 
for investigating the topological vacuum structure. This search was biased in favor 
of detecting instantons. Results concerning the ``instanton structure'' have been 
presented and discussed in the ``Confinement and Topology'' sessions at the annual
Lattice conferences until 2000~\cite{Negele:1998ev,Teper:1999wp, 
GarciaPerez:2000hq}. Beginning from 1998, for non-zero temperature calorons with
non-trivial holonomy and their dyon constituents~\cite{Kraan:1998pm,
Kraan:1998sn,Lee:1998bb,Bruckmann:2002vy,Bruckmann:2004nu}
have attracted more and more interest, although within a relatively small
community~\cite{Ilgenfritz:2002qs,Gattringer:2002wh,Gattringer:2002tg,
Diakonov:2002fq,Gattringer:2003uq,Diakonov:2004jn,Bruckmann:2004ib,
Ilgenfritz:2004zz,Ilgenfritz:2005um,Ilgenfritz:2006ju,Bornyakov:2007fm,
Bornyakov:2007xe,Bornyakov:2008im,Bruckmann:2009pa}.

There was always the hope to elucidate the different phases, both of pure
Yang-Mills theory and of full QCD, in terms of the topological structure.
The local behavior of the laminar sheets (and the resulting two-point
function seen with few cooling steps~\cite{Ilgenfritz:2008ia,Bruckmann:2011ve}),
however turned out to be not critically dependent on the phase.
Only in the IR, at best, the characteristic differences may become
visible~\cite{Moran:2008xq,deForcrand:2006my}.
One prominent example is the space-time anisotropy of the susceptibility in
slab-like subvolumes~\cite{deForcrand:1998ng} originally predicted in the
instanton-antiinstanton molecule model~\cite{Ilgenfritz:1994nt}.

In this paper we will use a specific (overimproved) kind of
cooling~\cite{GarciaPerez:1993ki,Bruckmann:2004ib}, however in the
Cabibbo-Marinari mode for $SU(3)$, and we are going to use it far beyond the
point where the IR structure (as a mixture of instantons and antiinstantons)
usually has been studied. Actually, this kind of action (see \Eq{eq:ecool} below)
was invented~\cite{GarciaPerez:1993ki} exactly for stabilizing
instantons or sphalerons~\cite{GarciaPerez:1993ki,GarciaPerez:1994kz}
under cooling. A similar purpose has been pursued in
Ref.~\cite{Bruckmann:2004ib} in order to study the elusive instanton
constituents at $T=0$. 

We will concentrate on the non-zero temperature case. We employ 
overimproved cooling in order to see in as far -- after eliminating 
all short-range fluctuations -- the emerging topological objects 
(``multi-caloron'' configurations) will clearly distinguish between 
the different phases or thermodynamic states in the neighbourhood
of the deconfinement transition for gluodynamics and crossover for
two-flavor QCD, respectively. The nature of these topological configurations 
in the case of $SU(3)$ gauge theory has been carefully considered already in
Ref.~\cite{Ilgenfritz:2005um}, and earlier for $SU(2)$ gauge theory in
Ref.~\cite{Ilgenfritz:2002qs}.

In this stadium of cooling the fields are either selfdual or antiselfdual
or trivial throughout the lattice, i.e. cooling has been employed until
it reaches the scale set by the whole lattice. Hence no more details than the 
total topological charge $Q$ can be read off and be associated to the original, 
thermalized configuration. This identification is convincing, however, since 
the topological charge mostly stabilizes after a few cooling steps.
No further localization~\cite{Koma:2005sw,deForcrand:2006my} of topological
charge characterizing the original Monte Carlo configurations is possible
in the final stadium of cooling.

Our study instead focuses on the following questions:
Under which circumstances nontrivial topological structures are stabilized
with respect to cooling ? How does this depend on the confining / deconfining
nature of the ensemble the gauge field configurations are taken from ?
What is the influence of the average Polyakov loop on the result of cooling ?

Our previous caloron studies~\cite{Ilgenfritz:2002qs,Ilgenfritz:2005um},
starting from thermalized (Monte Carlo) lattices at strong coupling or deep
in the confinement phase, did not consider the role of temperature in
detail. Next, at an intermediate scale of resolution, 
the study of topological objects in $SU(2)$ lattice fields at non-zero 
temperature~\cite{Ilgenfritz:2006ju,Bornyakov:2007fm,Bornyakov:2008im} 
has shown the profound difference between the confinement and deconfinement phase.
To be concrete, in this analysis either 50 APE smearing steps with a smearing 
parameter 0.45~\cite{Ilgenfritz:2006ju} or an overlap fermion analysis 
based on 20 lowest modes~\cite{Bornyakov:2007fm,Bornyakov:2008im} have been
applied. 

The following picture of the topological content of $SU(2)$ lattice gauge 
theory has emerged. At low temperatures one finds topological objects represented 
by non-dissociated calorons with maximally nontrivial 
holonomy~\cite{Kraan:1998pm,Kraan:1998sn,Lee:1998bb}. 
With increasing temperature~\cite{Ilgenfritz:2004ws} their composite nature
in the form of monopoles becomes recognizable. They start to dissociate into 
dyons of topological charge $\pm 1/2$ that appear (in the limiting case) as 
static $U(1)$ monopoles in the maximally Abelian gauge. 
Approaching the transition temperature $T_c$
(to the deconfining phase) from below, approximately one half of the calorons
were observed dissociated, retaining the symmetry between the constituent dyons.

Above the transition temperature, a non-zero expectation value of the averaged
Polyakov loop develops, which induces an asymmetry between the constituent dyons. 
This can be comprehensively explained in terms of the peak values of the local 
Polyakov loop~\cite{Bornyakov:2008im} : ``light'' (anti)dyons, with the local 
Polyakov loop of same sign as the averaged (prevailing background) Polyakov loop, 
become the most abundant topological objects, while ``heavy'' dyons or antidyons 
with the local Polyakov loop opposite to the average are 
suppressed~\cite{Bornyakov:2008bg,
Bruckmann:2009ne}.
The former ``heavy'' dyons may still carry highly localized fermionic 
zero modes bound to the ``defects'' detectable in the local Polyakov 
loop~\cite{Bruckmann:2011cc,Kovacs:2011tj}.
Nondissociated calorons are even more suppressed.

First steps towards the statistical mechanics of selfdual
dyons for the case of $SU(2)$ gauge theory, supposed to be
valid in the region around the critical temperature and
derived from the semiclassical partition function, have
been made in the papers~\cite{Shuryak:2012aa,Faccioli:2013ja}.

The present paper does not attempt to consider topology at the above mentioned 
intermediate scale of resolution. Concerning $SU(3)$ gauge theory,
this is left to an overlap fermion analysis of gauge field configurations
which is in progress.
The main outcome of the present study will be that the phase transition
between confinement and deconfinement in $SU(3)$ gluodynamics can also be 
characterized by the sharp change in the appearance of topological objects
remaining once cooling has hit the lattice scale. We confront this with 
corresponding (softer) results for the crossover known to replace the 
phase transition in the case of full QCD with $N_f=2$ dynamical fermions.

We claim that our observations concerning the behavior of action 
and Polyakov loop can be explained by the dyonic structure of calorons.
We investigate the volume and discretization effects for the 
topological susceptibility calculated by the use of topological charges
measured either at plateaus of topological charge or identified by the 
coincidence between $|Q|$ and $S/S_{\rm inst}$ and compare it with the topological
susceptibility of uncooled Monte Carlo configurations calculated by other
methods~\cite{Alles:1996nm,Gattringer:2002mr,Weinberg:2007tg}.

The following \Sec{sec:definitions2} contains the main definitions as well as 
some details of the simulations we have used here or where the full QCD 
configurations are taken from. \Sec{sec:results2} shows examples of cooling 
histories for $SU(3)$ gluodynamics in the confined and deconfined phases. 
We comment on the influence of the overimprovement parameter. The temperature 
dependence of the calculated topological susceptibility, both for pure gauge 
theory and for QCD, is discussed in \Sec{sec:results1}. In \Sec{sec:results3} 
the evolution of the averaged Polyakov loop during the cooling process for 
$SU(3)$ gluodynamics and for full QCD, both in the confined and deconfined phases, 
is presented and the interpretation with the help of the dyonic picture of the 
gauge field ensemble is discussed. \Sec{sec:conclusions} is reserved for 
conclusions and discussion. In the Appendix we recall some facts on $SU(3)$ 
calorons and dyons.

\section{Thermal ensembles}
\label{sec:definitions2}

For the study of the phase transition in the $SU(3)$
pure gauge theory we employ the standard Wilson action $S_W$ with the lattice 
coupling $\beta = 6/g_0^2 \,$ where $~g_0$ is the bare coupling constant.
To determine the corresponding lattice spacing $a$ as a function of $\beta$, 
for this action we have used the Necco--Sommer parametrization \cite{Necco:2001gh}. 
In what follows we will refer to this case as {\it gluodynamics}.

To study the  topological aspects of the phase transition or crossover
in a theory with dynamical quarks we have studied gauge field configurations
generated with the gauge action $S_W$ and $N_f=2$ dynamical flavors of
nonperturbatively $O(a)$ improved Wilson fermions (clover fermions). The 
configurations had been produced by the DIK
collaboration~\cite{Bornyakov:2004ii} 
using the Berlin QCD code (BQCD)~\cite{Nakamura:2010qh}. The improvement 
coefficient $c_{SW}$ was determined nonperturbatively \cite{Jansen:1998mx}.
The lattice spacing and pion mass has been determined by interpolation of $T=0$
results obtained by QCDSF \cite{Gockeler:2006jt}. In the
remainder of the paper this case will be referred to as {\it full QCD}.

\vspace{2mm}

Our calculations were performed on asymmetric lattices 
with the four-dimensional volume $V= a^4 L_t\cdot L_s^3$,
where $L_t$ is the number of sites in the
time ($4th$) direction. The temperature $T$ is given by $~T=1/aL_t~$.
In the case of gluodynamics
we have employed $L_t=4$, $L_s=16$ as well as  $L_s=24$ lattices, 
for which we have generated and analyzed 1000 and 500 configurations, 
respectively, at a set of $\beta$ values. The coupling 
$\beta_c = 5.692$ \cite{Iwasaki:1992ik} 
characterizes the transition at $L_t=4$. 
It  corresponds to the critical temperature
$T_c \simeq 300$ MeV \cite{Gattringer:2002mr}. 
In order to study finite lattice spacing effects we have
also simulated  $L_t=6$, $L_s=24$ lattices with statistics of 
500 gauge field configurations for each  $\beta$ value.
The phase transition in this case takes place at 
$\beta_c = 5.894$ \cite{Iwasaki:1992ik}.
In order to keep autocorrelations small, all
measurements were made on configurations separated by 500 sweeps.

In the case of full QCD~\cite{Bornyakov:2004ii} we have analysed 
configurations produced on lattices with
$L_t=8$ and spatial sizes $L_s=16$ (500 configurations at each
$T/T_c$) and $L_s=24$ (200 configurations at each  $T/T_c$).
The temperature $T$ was effectively varied at fixed $\beta$-value
by changing the Wilson fermion hopping parameter $\kappa$, 
i.e. the quark mass was not kept constant.
The chiral crossover temperature $T_c \approx 230$ MeV 
was determined in Refs. \cite{Bornyakov:2004ii,Bornyakov:2011jm}
at a pion mass value of $O(1~\mathrm{GeV})$. 

\section{Cooling histories}
\label{sec:results2}

Technically, we cool down each gauge field configuration by means of the usual
Cabibbo-Marinari cooling procedure, now with respect to an overimproved
action~\cite{GarciaPerez:1993ki} that has been used
before~\cite{Ilgenfritz:2005um} for the study of $SU(3)$ calorons and multicalorons.
We monitor the cooling process and search for plateaus of the action and topological 
charge appearing in the cooling history. We consider
\begin{itemize}
\item either the first plateau of the topological charge, when the
topological charge $Q$ stays near some integer value $n$ ($|Q - n|<0.1$)
for at least 100 cooling sweeps,
\item or the first (anti)selfdual plateau, when the difference between
$|Q|$ and the action $S$ (measured in units of instanton action $S_{inst}$)
becomes small ($S/S_{inst}-|Q|<0.1$).
\end{itemize}

For both gluodynamics and full QCD
in the confined phase  we observe the two kinds of plateaus
to coincide, whereas
deep in the deconfined phase no plateaus of any kind are seen.
In the transition region (just above the transition
to the deconfined phase or for full QCD above the crossover)
the first topological charge plateaus can be
different from the first (anti)selfdual plateaus.
Moreover, both plateaus are unstable.

We use overimproved cooling \cite{GarciaPerez:1993ki} because it partially
stabilizes calorons with respect to shrinking and falling ``through the
meshes of the lattice''.  The following parametrization of the lattice action
($\linkhmu\equiv U_\mu(x) \in SU(N)$ link variable) allows for over- and
underimprovement~\cite{GarciaPerez:1993ki},
\beqa
&S(\veps)=
\sum_{x,\mu,\nu}\frac{4-\veps}{3}\,Re\, \Tr\left(1-\plaq\right)\\\nonumber
&+\sum_{x,\mu,\nu}\frac{\veps-1}{48}\,Re\, \Tr\left(1-\twoplaq\right).
\label{eq:ecool}
\eeqa
Expanding in powers of the lattice spacing $a$ one finds~\cite{GarciaPerez:1993ki},
\begin{eqnarray}
S(\veps) &= & \sum_{x,\mu,\nu}a^4\Tr\left[-\frac{1}{2}F_{\mu\nu}^2(x)+\frac{\veps
a^2}{12}(\cD_\mu F_{\mu\nu}(x))^2 \right] \nonumber \\
&& + \cO(a^8)
\label{eq:epsac2}
\end{eqnarray}
(note that no implicit summation convention is implied in this formula).
$S(\veps~=~1)$ corresponds to the Wilson action, and the sign of the
leading lattice artifacts is simply reversed by changing the sign of
$\veps$. Based on a discretized continuum one-instanton solution of
size $\rho$, one finds 
\begin{equation}
S(\veps)=8 \pi^2 \ \left[ 1-
\frac{\veps}{5}(a/\rho)^2+\cO(a/\rho)^4 \right] \; , 
\end{equation}
suggesting that
under cooling $\rho$ will decrease for $\veps>0$ and increase for $\veps<0$.

\begin{figure*}[!htb]
\centering
\includegraphics[width=.4\textwidth]{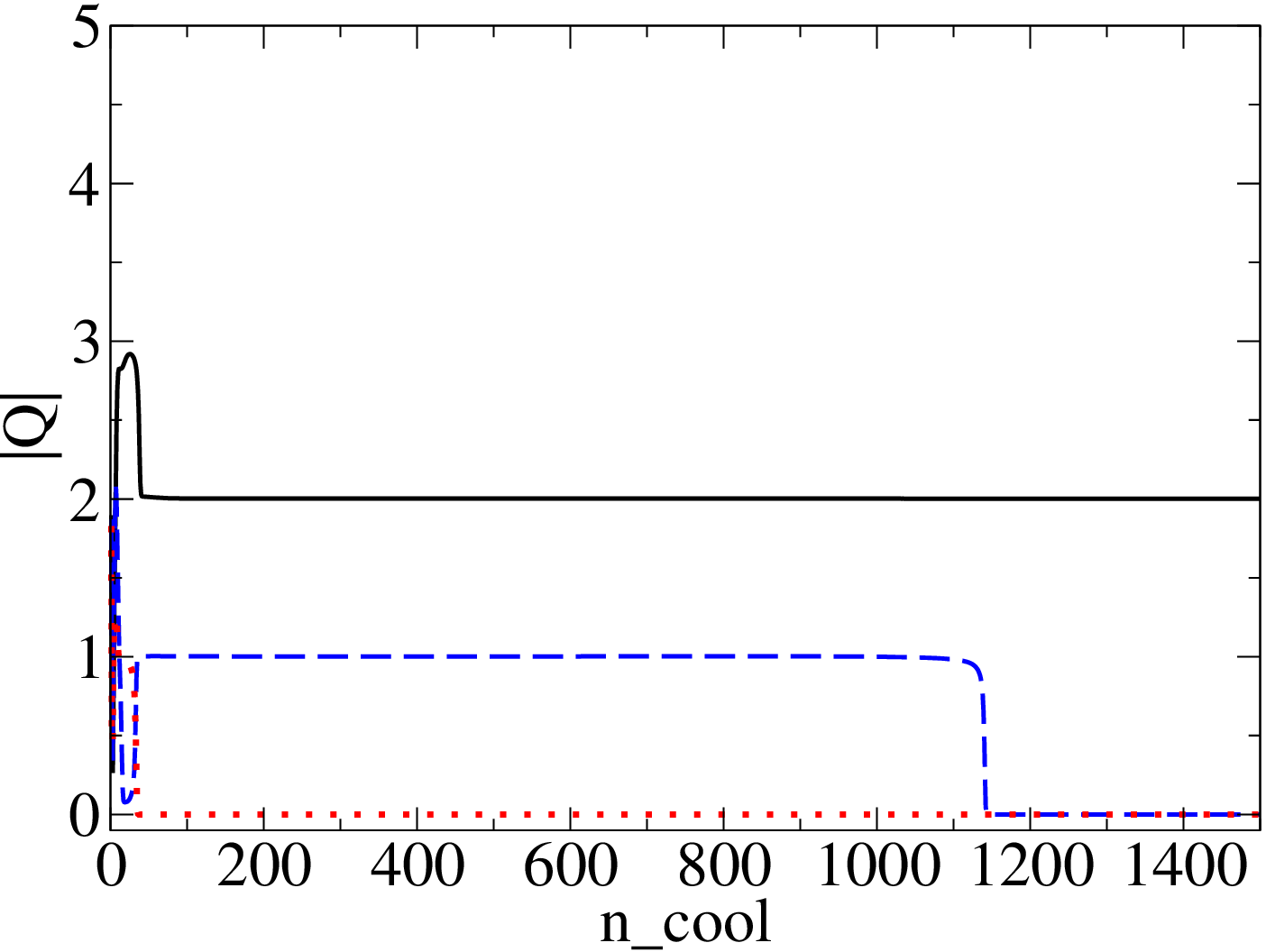}
\hspace{1.5cm}
\includegraphics[width=.4\textwidth]{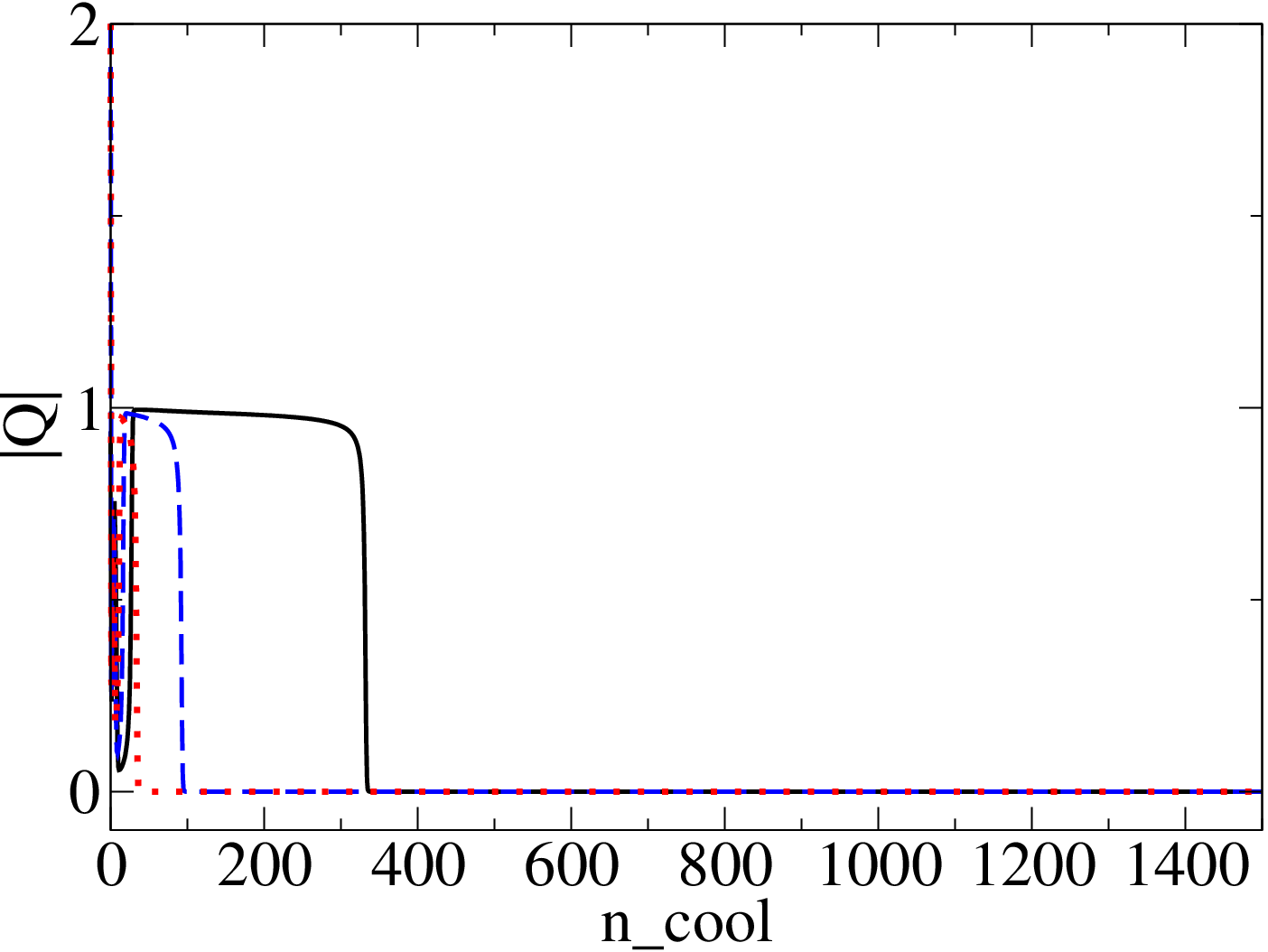}
\caption{Cooling histories of $|Q|$ corresponding to $\veps=1$
(Wilson action, red dotted line), $\veps=0$ (slightly overimproved action,
blue dashed line) and $\veps=-1$ (overimproved action as used in the rest
of this paper, black solid line). Left: different cooling histories for one
configuration from the confined phase. Right: different cooling histories
for a (conditionally stable) caloron in a configuration taken from the
transition region.}
\vspace{2cm}
\label{fig:epscomparison}
\end{figure*}

In order to illustrate the influence of the parameter $\veps$, we present
in Fig. \ref{fig:epscomparison} three cooling histories for $\veps=1$
(Wilson action), $\veps=0$ (slightly overimproved action), and $\veps=-1$
(strongly overimproved action, our choice in this paper). The left panel
shows this comparison for a gluodynamics configuration taken from the
confined phase, the right panel for a gluodynamics configuration from the
transition region.
In the full statistics analysis of this paper we use $\veps=-1$.

Both action and topological charge were calculated with the help of the 
improved lattice version of the
field strength tensor \cite{BilsonThompson:2001ca} as
$S/S_{\rm inst}
= \sum_{x,a} \left( E^a \cdot E^a + B^a \cdot B^a \right)/(16 \pi^2)$
and $Q = \sum_{x,a} E^a \cdot B^a/(8 \pi^2)$.

\begin{figure*}[!htb]
\centering
\includegraphics[width=.4\textwidth]{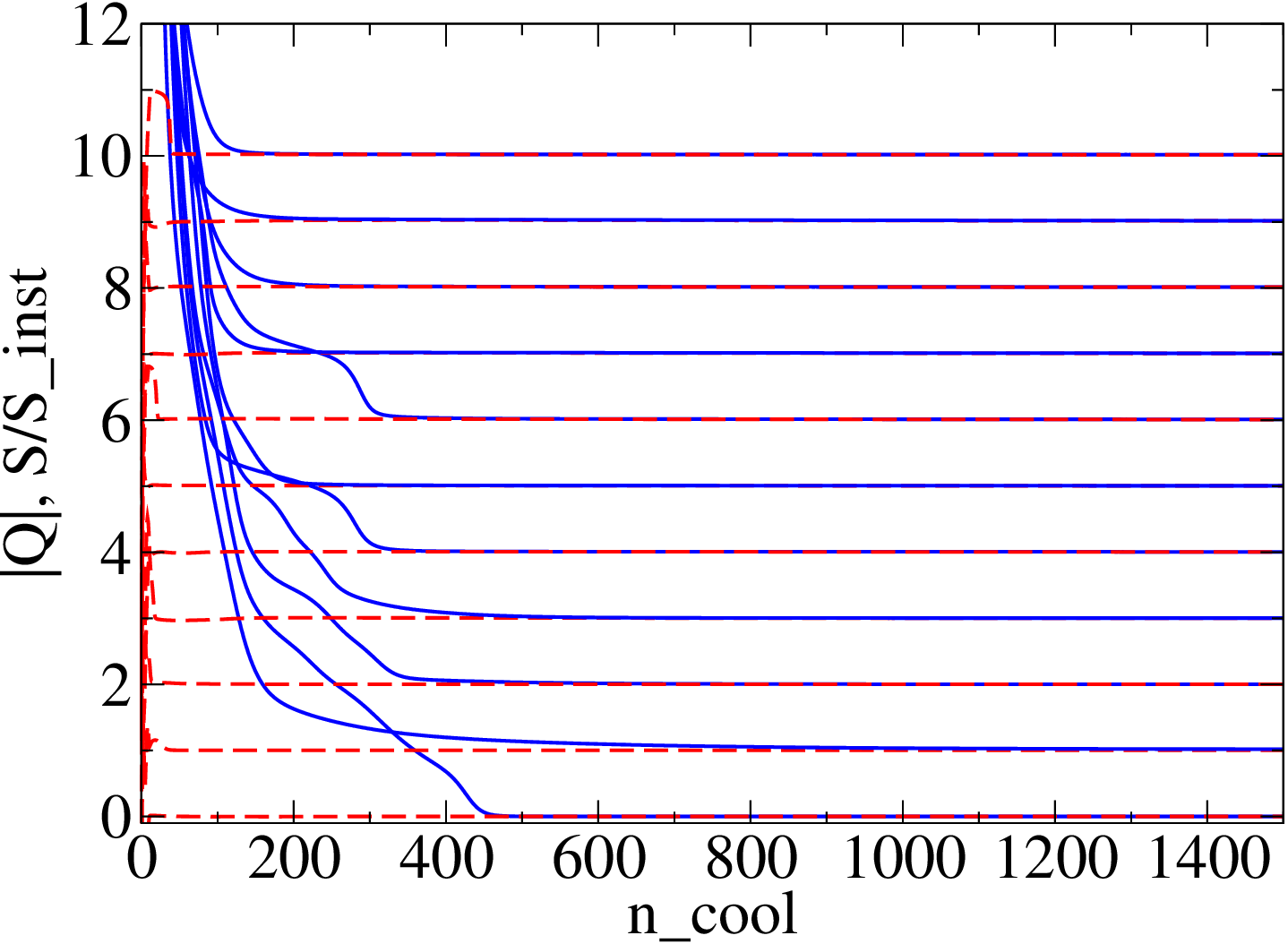}
\hspace{1.5cm}
\includegraphics[width=.4\textwidth]{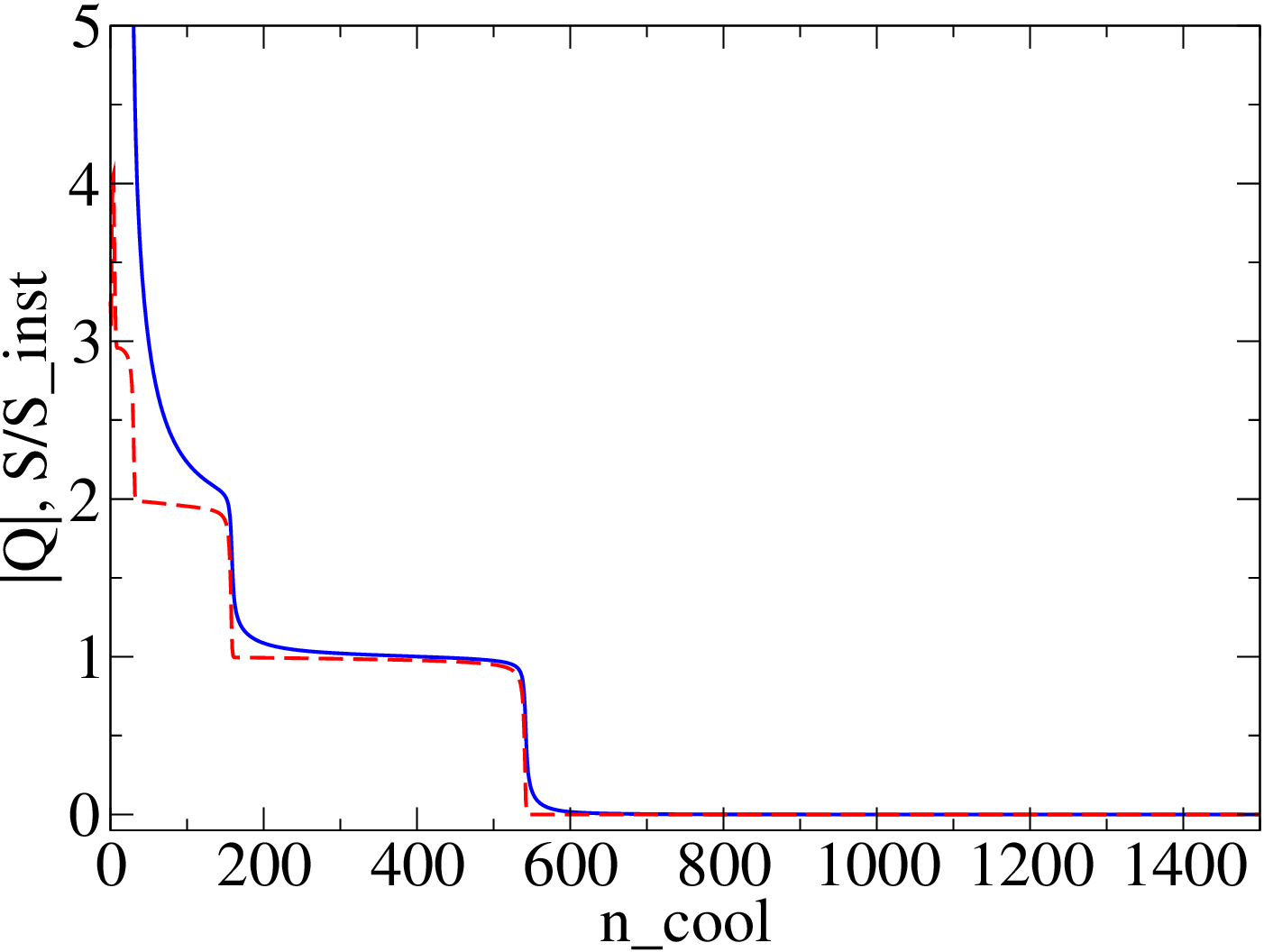}
\caption{Typical cooling histories of configurations from the confined phase
(left). The cooling history of an (finally unstable) caloron in a configuration
taken from the transition region (right). The red dashed lines represent $|Q|$,
the blue solid lines represent $S$.}
\vspace{2cm}
\label{fig:coolingcondec}
\end{figure*}

As a result, for gluodynamics in the confined phase we find 
calorons to be stable during extended cooling.
This is well seen in the left part of 
Fig.~\ref{fig:coolingcondec}. One notices that the modulus of the 
topological charge approaches some plateau very early. 
The action (expressed in units of instanton action) later converges 
towards this plateau as well. 
The topological charge plateau is stable, what can be understood as stability 
of calorons, which (in the confined phase) are appearing in the form of three 
finite action dyons. 

For gluodynamics in the transition region (early deconfined phase), 
calorons which are rarely present in lattice configurations are unstable 
and cascading down to a trivial 
vacuum as it is seen in the right part of Fig. \ref{fig:coolingcondec}.
The first topological charge plateau sometimes happens to be different
from the first (anti)selfdual plateau (this is the case shown in the
right panel of Fig. \ref{fig:coolingcondec}). In 
this temperature range the topological susceptibility
calculated by the use of topological charges measured on the two kinds
of plateaus takes different values, but the difference is never larger than 6\%. 

Deep in the deconfined phase, at $T/T_c = 1.8$,
we have found that both topological charge and action go to zero values
without any intermediate plateaus.
 
\begin{figure}
\vspace{1cm}
\includegraphics[width=0.45\textwidth]{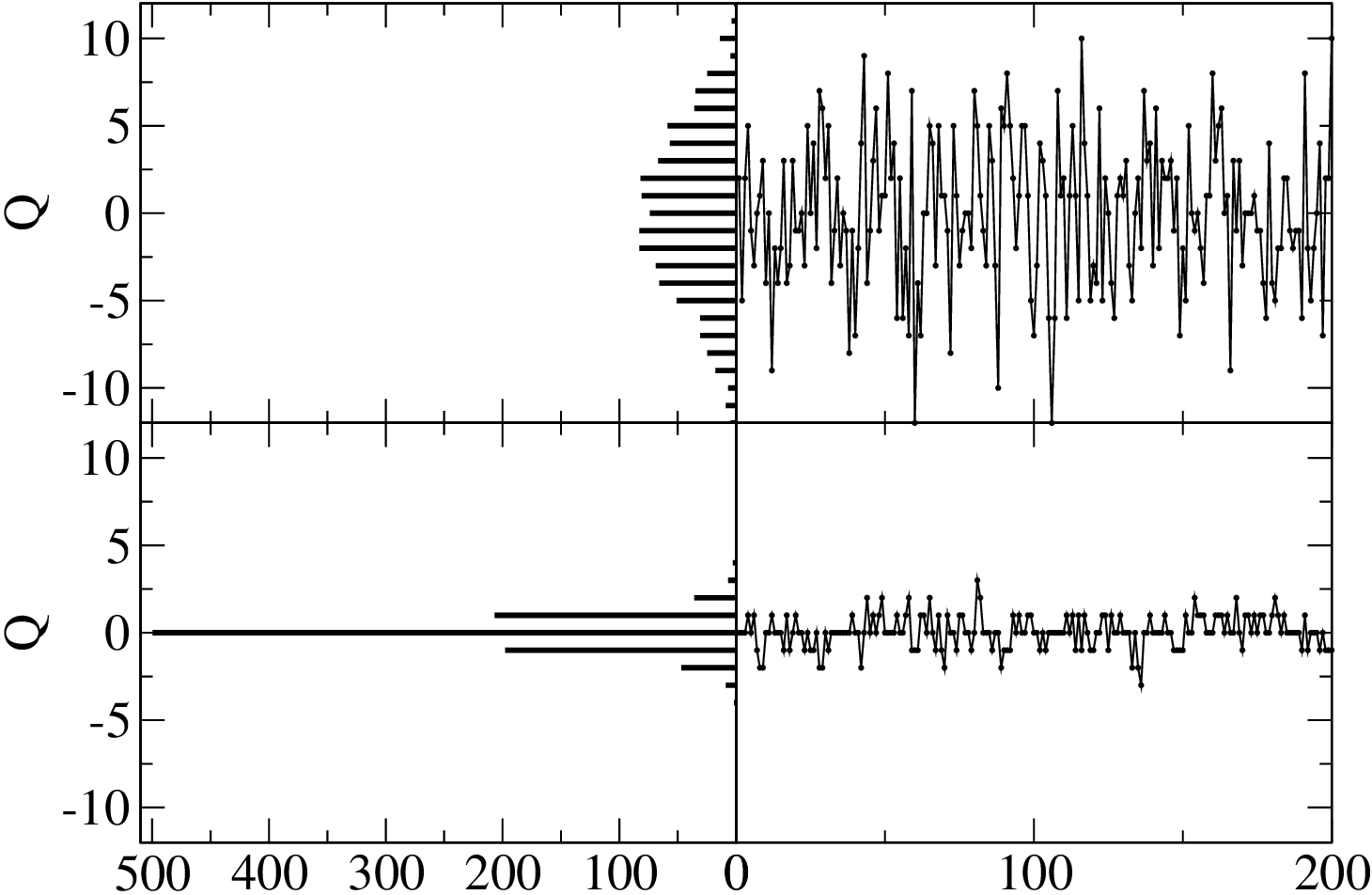}
\caption{The distributions of topological charges
(left) and the Monte Carlo time histories along the Monte Carlo chain (right)
of pure gluodynamics are shown, for the confined phase in the upper part and
for the deconfined phase in the lower part. Confinement is represented by a
sample at $T=0.883~T_c$, deconfinement is illustrated by a sample
at $T=1.117~T_c$, in both cases with an actual statistics of 1000
$L_t=4$,$L_s=16$ lattice configurations (only partly shown in the history). }
\label{fig:distr}
\end{figure}

In what follows we will use only topological charge values
determined on the first topological charge plateau. 
In Fig. \ref{fig:distr} again for gluodynamics
the Monte Carlo time histories of the topological charge are shown for the first 
200 measurements and the histogram of these values is presented with full statistics.
These results are presented for the confined phase ($T=0.883~T_c$)
in the upper part and for the deconfined phase ($T=1.117~T_c$) in the
lower part of the figure. The data indicate that the topological charges are
well decorrelated.

We found very similar features of cooling plateaus in full QCD.

\section{Temperature dependence of the topological susceptibility}
\label{sec:results1}

\begin{figure*}
\vspace{1 cm}
\includegraphics[width=.45\textwidth]{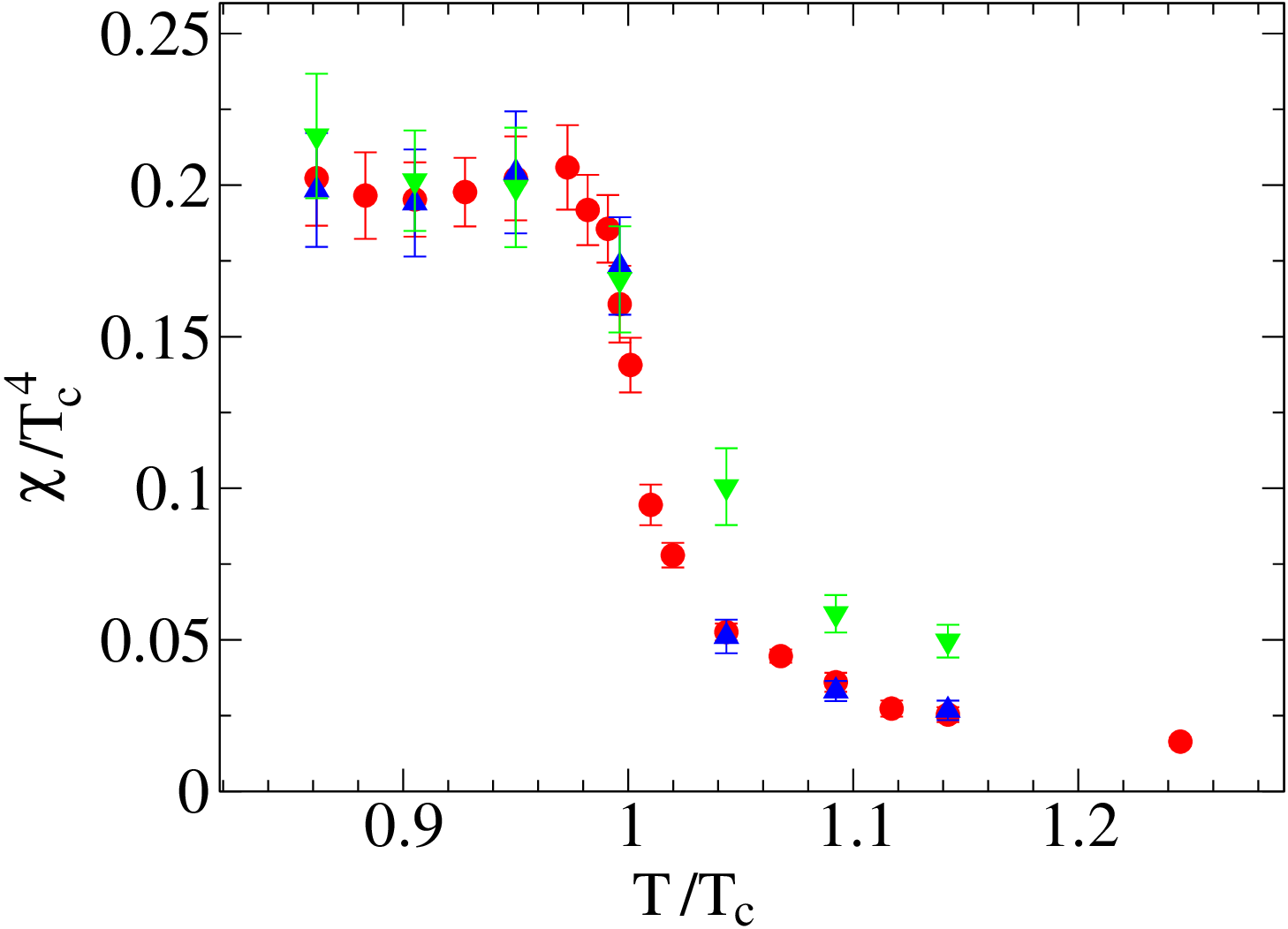}%
\hspace{1 cm}
\includegraphics[width=.45\textwidth]{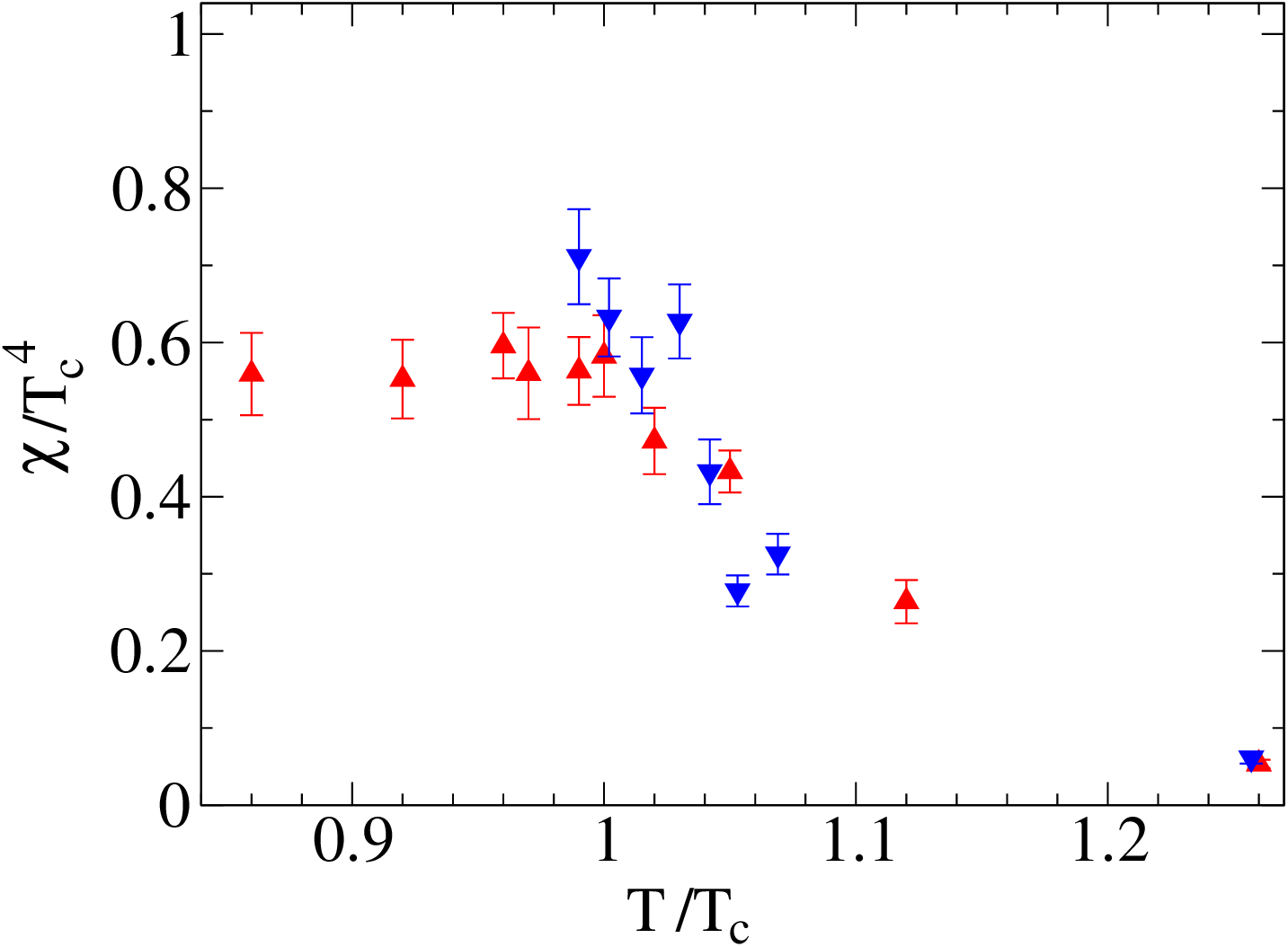}\\
\caption{Left: topological susceptibility for  gluodynamics
on the lattices $16^3*4$ (red circles), $24^3*4$ (blue up triangles) and
$24^3*6$ (green down triangles).
Right: susceptibility for full QCD on the lattices $16^3*8$ (red up
triangles) and $24^3*8$ (blue down triangles).  }
\label{fig:phtr}
\end{figure*}

Our results for the topological susceptibility
\beq
\chi=\frac{1}{V} \langle Q^2 \rangle
\label{eq:topsuscept}
\eeq
for gluodynamics are presented in the left part of Fig. \ref{fig:phtr} for
lattices $16^3 \cdot 4$ (red circles), $24^3 \cdot 4$ (blue up triangles) 
and $24^3 \cdot 6$ (green down triangles). Respective results for full 
QCD are shown in the right part of Fig. \ref{fig:phtr}.
They were obtained by applying cooling to the lattice ensembles of the DIK
collaboration for $16^3 \cdot 8$ (red up triangles) and for $24^3 \cdot 8$ 
(blue down triangles).

From the left part of Fig.~\ref{fig:phtr} we see that the results obtained
on lattices $16^3 \cdot 4$ and $24^3 \cdot 4$ are close to each other.
This means  that finite volume effects are practically absent.
The comparison of results from lattices $16^3 \cdot 4$ and $24^3 \cdot 6$ shows
that discretization effects in the confinement phase are small while
sizeable discretization effects are present in the deconfined phase.
In section \ref{sec:results3} we suggest an explanation of this feature.
On the right panel of Fig.~\ref{fig:phtr} -- devoted to full QCD -- 
one can see that $\chi$ is more or less constant below $T_c$ and starts to 
decrease slowly above $T_c$. The data also indicates finite volume 
effects to be small, similarly to the gluodynamics case.

\begin{figure}
\vspace{1 cm}
\includegraphics[width=.4\textwidth]{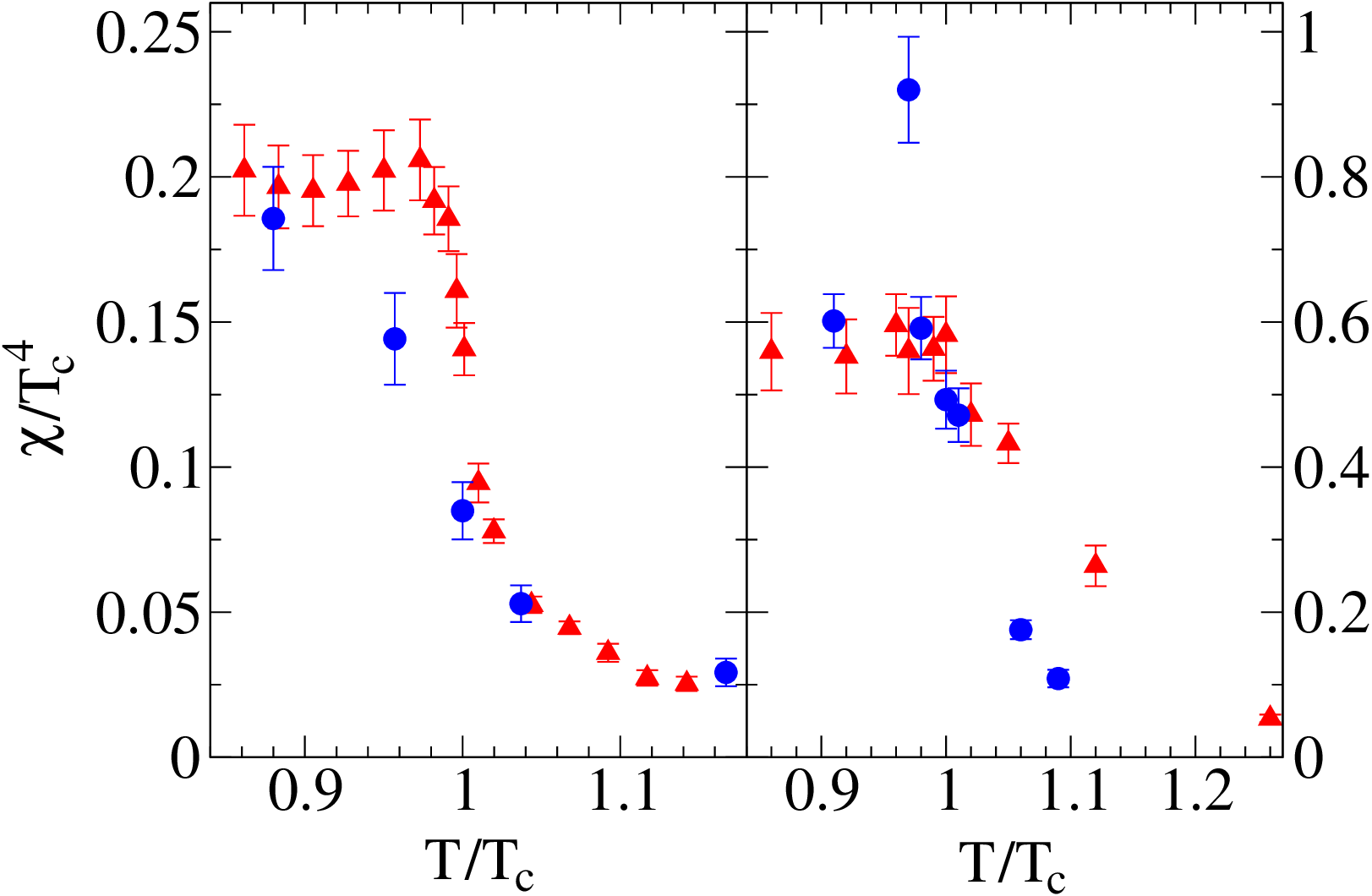}\\
\caption{Left: comparison of the topological susceptibility for
gluodynamics obtained in Ref.~\cite{Gattringer:2002mr} (blue circles) with
the topological susceptibility calculated in the present paper (red up
triangles).
Right: comparison of topological susceptibility for full QCD from
Ref.~\cite{Weinberg:2007tg} (blue circles) with the topological susceptibility
calculated in the present paper (red triangles).}
\label{fig:comp}
\end{figure}
\begin{figure}
\vspace{1 cm}
\includegraphics[width=.4\textwidth]{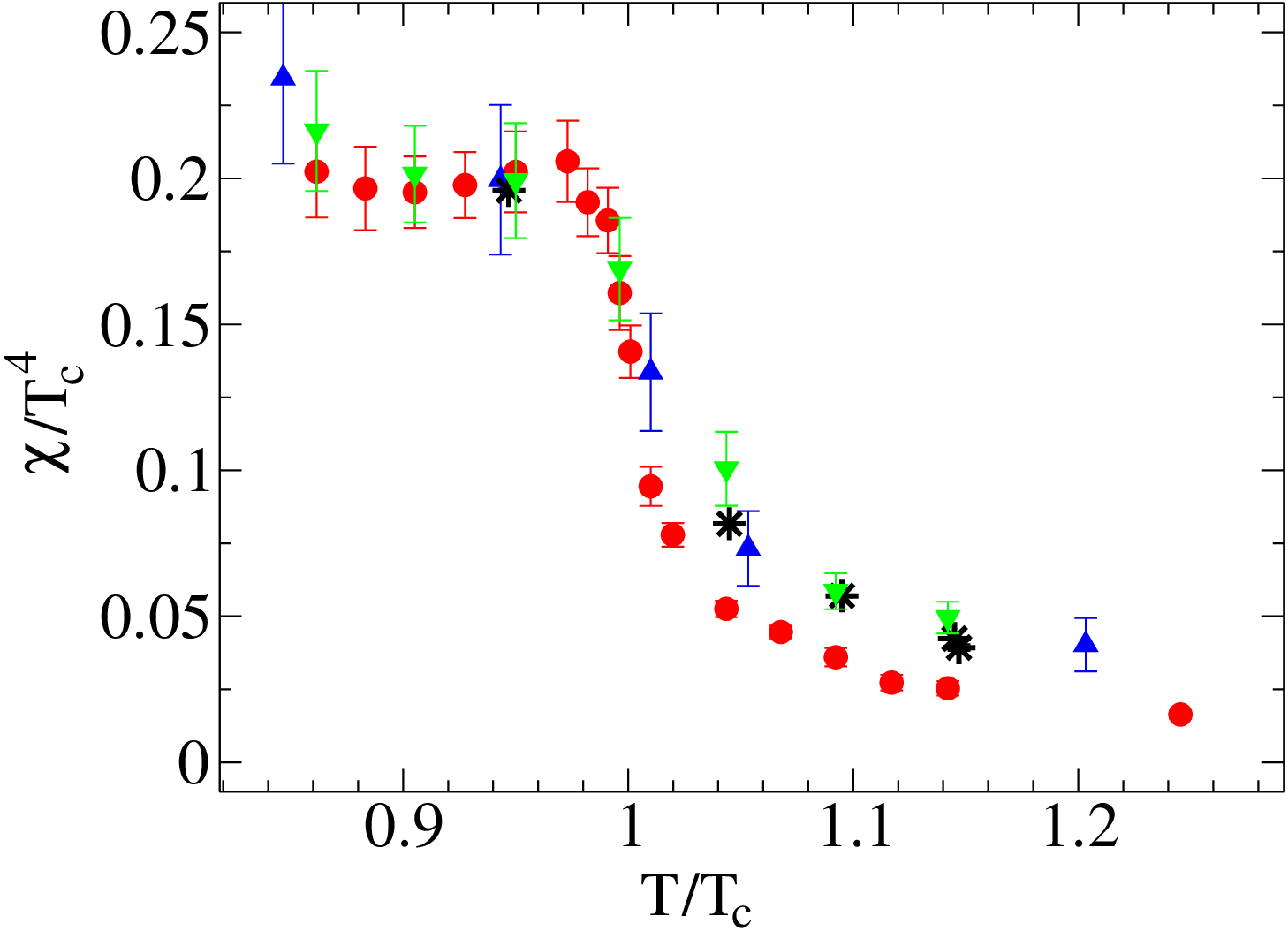}\\
\caption{Comparison of the topological susceptibility for  gluodynamics
from Ref.~\cite{Alles:1996nm} (blue up triangles) and from
Ref.~\cite{Bonati:2013tt} (black stars) with the topological susceptibility
calculated in the present paper on the lattices $24^3*6$ (green down triangles)
and $16^3*4$ (red circles).}
\label{fig:compgia}
\end{figure}

The topological susceptibility calculated in this work for gluodynamics
is compared in Fig.~\ref{fig:comp} (left) with the topological susceptibility 
calculated by the use of the index theorem for a chirally improved Dirac operator 
without cooling~\cite{Gattringer:2002mr}.
For full QCD the comparison with results obtained by the use of the
index theorem for the overlap Dirac operator~\cite{Weinberg:2007tg}
is presented in Fig.~\ref{fig:comp} (right).
In the case of gluodynamics we find rather good agreement for all data points
(uncertainties in the determination of $T/T_c$ should be taken into account)
apart from one point at $T/T_c \approx 0.95$.  The most essential qualitative
difference between our results and those
of Ref.~\cite{Gattringer:2002mr} is that our data show a sharp drop of $\chi$ 
at the transition while the data of Ref.~\cite{Gattringer:2002mr} indicate a 
much smoother behavior. For full  QCD the results agree in the confinement phase 
but disagree in the deconfinement phase.

In Fig.~\ref{fig:compgia} we compare our gluodynamics results with the early 
results of Ref.~\cite{Alles:1996nm} (blue up triangles). In this paper the field
theoretical method was applied for measuring the two-point function of
the topological density. This method circumvents the problem of defining
a topological charge for each configuration and uses multiplicative and
additive renormalization in order to relate the two-point function at
zero momentum to the continuum topological susceptibility.
The comparison is made between our results for lattices $24^3 \cdot 6$ (green down
triangles) and $16^3 \cdot 4$ (red circles) and the results of
Ref.~\cite{Alles:1996nm} which refer to a $32^3 \cdot 8$ lattice (blue up triangles). 
The agreement with our results obtained on finer $24^3 \cdot 6$ lattices looks very 
good in both phases. This could be an indication of the absence of 
discretization effects on our $L_t=6$ lattices.
This is supported by comparison with
very recent results of Ref.~\cite{Bonati:2013tt} where few sweeps of cooling
were used to evaluate the topological charge on a lattice of size $40^3 \cdot 10$ .

\section{Interpretation in terms of a dyonic picture}
\label{sec:results3}
Let us interpret our findings in terms of the caloron or dyon 
content of the gluonic field ensembles and of the behavior 
of the average Polyakov loop (related to the holonomy of the caloron 
configurations) during the cooling process. 

\begin{figure*}[!htb]
\centering
\includegraphics[width=.4\textwidth]{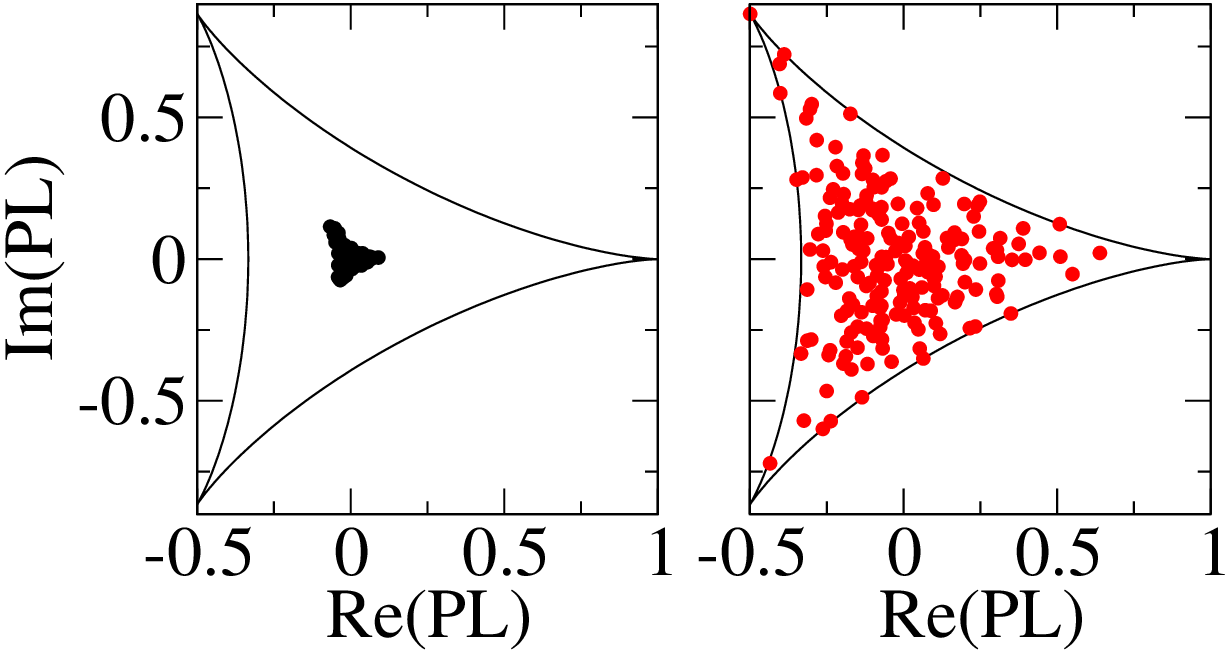}
\hspace{1.5cm}\includegraphics[width=.4\textwidth]{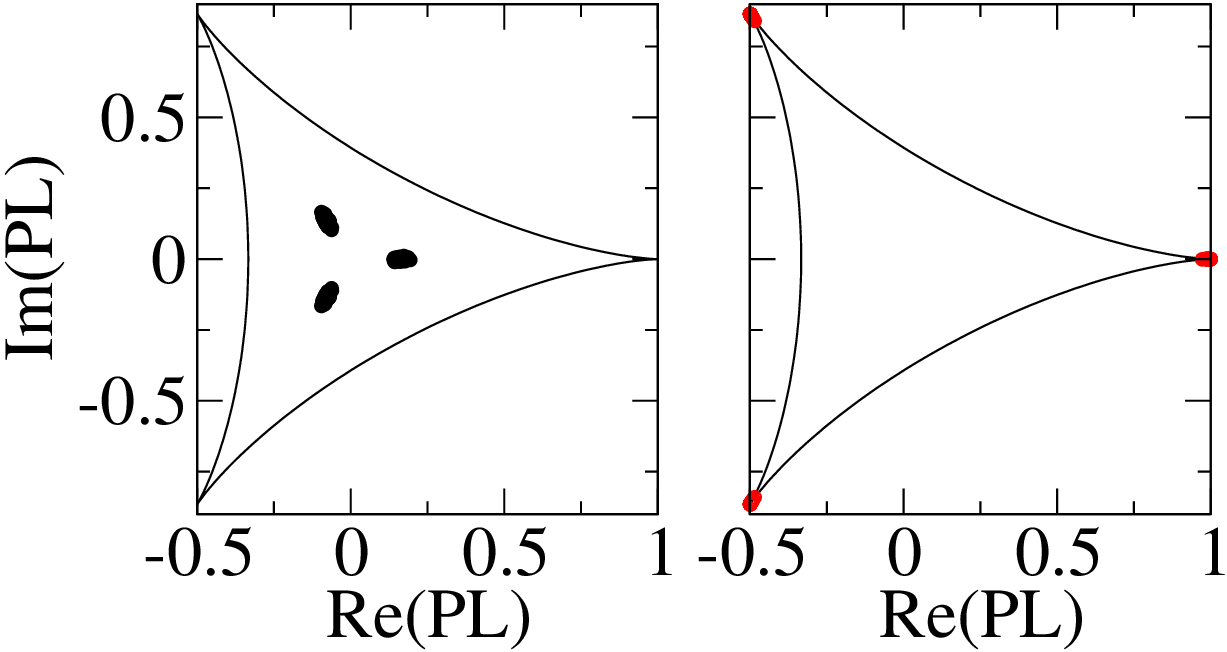}
\caption{
For gluodynamics, scatter plots of holonomy (spatially averaged Polyakov loop) 
for thermalized configurations and after cooling.
Left panels: for the confined phase ($T/T_c=0.97$) and at the corresponding
stable, non-trivial (anti)selfduality plateaus; 
right panels: for the deconfined phase ($T/T_c=1.09$) and correspondingly
after 1500 cooling sweeps at (almost) vanishing action values.
}
\label{fig:purepolscat}
\vspace{1.5cm}
\end{figure*}

We discuss the case of gluodynamics first.
The action of a caloron does not depend on the holonomy (see Appendix
and Fig.~\ref{fig:triangle} for an illustration of this fact).
So, while the holonomy changes considerably in the process of cooling,
the action of the (multi)caloron (gradually formed in the lattice field
configurations by cooling in the confined phase) is not changing.
Due to this stability of the (multi)caloron action  
the holonomy has no preferred
direction to evolve (see the left panel of Fig. \ref{fig:purepolscat}
referring to the confined phase before and after cooling).
In the deconfined phase on the other hand, even if the three center
sectors are equally represented (see the right panel of
Fig. \ref{fig:purepolscat} referring to the deconfined phase before and
after cooling), under cooling the holonomy always moves towards the
corresponding corner of the plot, since this is the direction where
asymmetric dyon-antidyon pairs according to the experience in the 
$SU(2)$ case (see the Introduction) could minimize their total action.

\begin{figure*}[!htb]
\centering
\includegraphics[width=.4\textwidth]{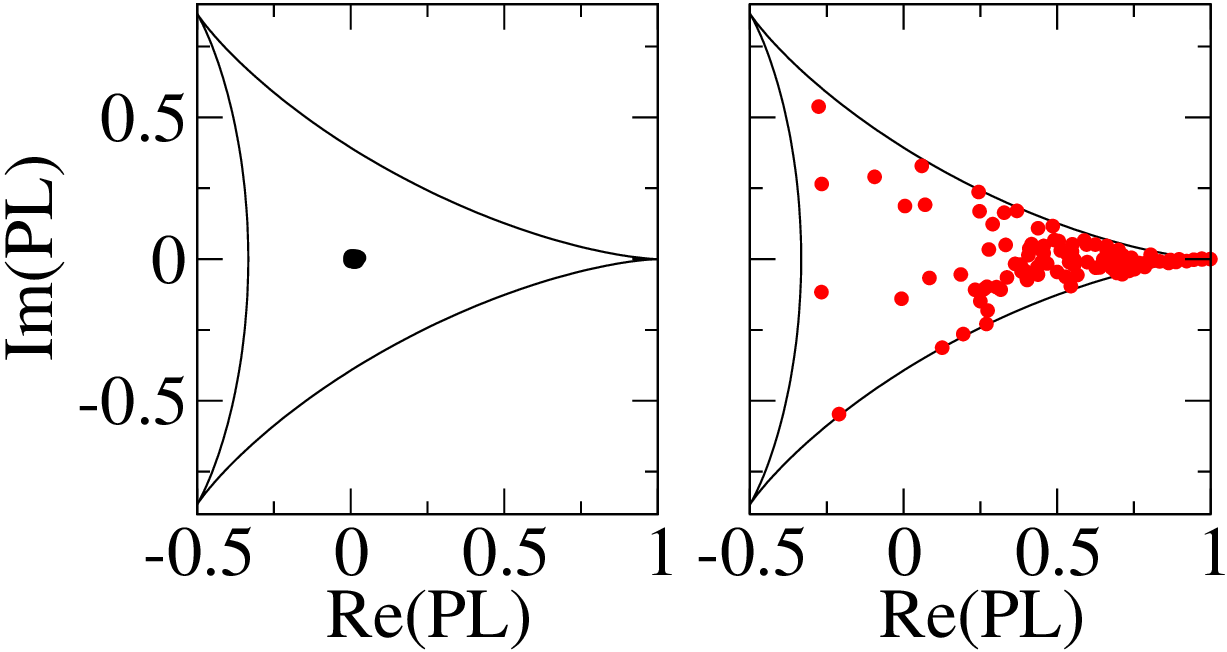}
\hspace{1.5cm}\includegraphics[width=.4\textwidth]{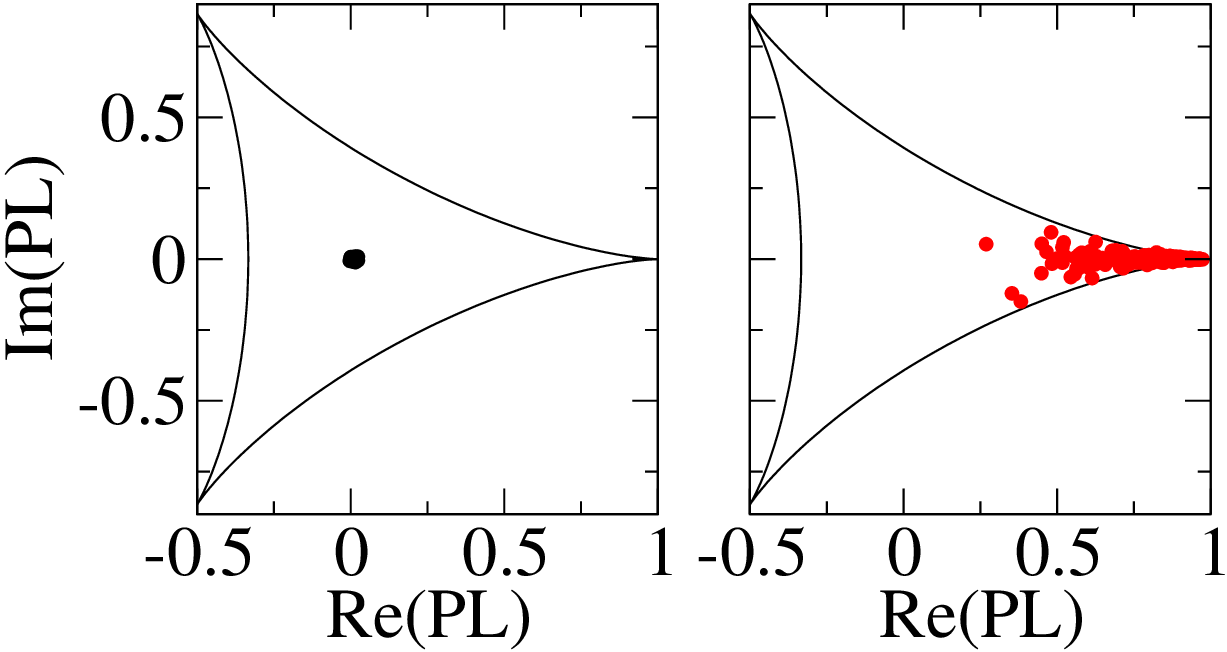}
\caption{
For full QCD, scatter plots of holonomy (spatially averaged Polyakov loop) 
for thermalized configurations and after cooling at stable, non-trivial 
(anti)selfduality plateaus.
Left panels: for the confined phase ($T/T_c=0.96$); 
right panels: for the deconfined phase ($T/T_c=1.05$).
}
\label{fig:QCDpolscat}
\end{figure*}

We compare this with full QCD. In this case, center symmetry
is slightly violated by the dynamical fermions already at
low temperature (``confined phase''). The left panel of Fig.~\ref{fig:QCDpolscat}
shows where the holonomy moves to in the result of cooling. We see how
the small direct violation of center symmetry in the confined phase of
full QCD is amplified by cooling (again due to the effect of minimization
of the action of asymmetric dyon-antidyon pairs). The right panel of 
Fig.~\ref{fig:QCDpolscat} shows the holonomies before and after cooling in the 
deconfined phase. Here direct and spontaneous violation of central symmetry
are summed and the asymmetry in the evolution of holonomy is more profound.

The stability of calorons in the confined phase and their partial instability
in the deconfined phase, both in the case of gluodynamics, can be understood 
if one takes into account that during the process of cooling the holonomy 
remains nontrivial in the confined phase whereas it rapidly becomes trivial 
in the deconfined phase.
In the confined phase the caloron, for nontrivial holonomy consisting
out of three localized dyons, should first form its nondissociated state by
recombination of three dyons before it could ``drop through the meshes''.
This, however, is improbable in the process of overimproved cooling due to 
the repulsion between dyons~\cite{Bruckmann:2004ib} that the overimproved 
action induces.

In the deconfined phase of gluodynamics with holonomy being almost trivial
in the cooling process, two light, delocalized dyons forming a caloron 
together with a heavy, localized dyon are already overlapping with the latter.
This could be the reason of calorons being finally unstable in the deconfined 
phase.

In QCD, for comparison, the transition from stability to instability develops
not so sharply and this is connected to the less rapid change
with temperature of the evolution of the holonomy during cooling
as it can be seen in Fig. \ref{fig:QCDpolscat}.

As we have already seen, the comparison of results from lattices $16^3*4$
and $24^3*6$ shows the presence of sizeable discretization effects in the
deconfined phase. They can be interpreted as follows. We have discussed the 
partial instability of calorons in the deconfined phase.
It is quite natural to expect that calorons are more stable on a finer lattice
where more effort is needed to let them ``fall through the smaller meshes''
of the lattice. Hence, on finer lattices the topological susceptibility
obtained by cooling in the deconfined phase should be larger.

\section{Conclusions}
\label{sec:conclusions}

In this work we have studied numerically, across the phase transition in
gluodynamics and across the crossover in full  QCD, the topological
objects, that remain conserved or decay after a suitably chosen cooling procedure.
We have applied the cooling method to find the topological content of the
first topological charge plateau and of the first (anti)selfdual plateau in
the cooling history of lattice configurations. 

In gluodynamics, we have been able to recognize the phase
transition point $T_c$ as the point separating (within a rapid
transition) temperature regions with and without surviving (anti)selfdual
(topological) objects. We compare our observations for gluodynamics with full 
QCD with $N_f=2$ dynamical flavors in the vicinity of the crossover temperature 
$T_c$.

We are convinced that the (anti)selfdual plateaus can be fully characterized
by (multi)caloron configurations as discussed in ~\cite{Ilgenfritz:2005um}.
In the rare cases, in the transition region, where the first topological charge 
plateau could differ from (anti)selfdual plateaus it is natural to assume an 
admixture of caloron-anticaloron or dyon-antidyon pairs.

We calculated the topological susceptibility based on the
topological charges of individual lattice configurations identified in this
way. We investigated the transition or crossover, respectively,
from the confined to the deconfined phase with the help of this topological 
susceptibility $\chi$.

In pure gluodynamics with increasing temperature, $\chi$ turned out to drop 
sharply down from a value close to the zero-temperature value just
at the (first order) phase transition, the latter determined from the 
behavior of the Polyakov loop or other quantities. Our finding tells
us that the topological susceptibility determined via (overimproved) 
cooling can be used as an alternative indicator for the transition itself. 

At the same time the drop off of $\chi$ could be viewed in terms 
of stable (multi)calorons on the confinement side and in terms of unstable
dyon-antidyon pair or caloron configurations on the deconfinement side.
The different behavior on both sides could be explained as triggered by the 
holonomy directly related to the 3-space averaged Polyakov loop for each
gauge field configuration. We demonstrated that (over)improved cooling drives 
the holonomy to non-trivial (trivial) values in the confinement (deconfinement)
phase with the consequence of getting stable (unstable) (multi)caloron
configurations due to their symmetric (non-symmetric) dyon content.     

In full QCD the drop of the topological susceptibility turned out to be
softer and to set in at a temperature that is slightly above the pseudocritical 
temperature $T_c$ determined by the maximum of the Polyakov loop
susceptibility~\cite{Bornyakov:2004ii}.
Also in this case the former discussion of stable (below the crossover)
or unstable (above the crossover) (multi)caloron configurations applies,
since the holonomy behaves similarly in spite of the $Z(3)$ breaking 
effect by the dynamical fermion degrees of freedom taken into account.  

We compared our results on topological susceptibilities with those
of other authors employing different methods and have found a reasonable 
agreement.

All together, our results give us some confidence that large-scale 
topological objects play a major role in the change of thermal gauge
field ensembles at the deconfinement phase transition in gluodynamics
as well as at the crossover phenomenon in full QCD.

\subsection*{Acknowledgments}
B.V.M. appreciates the support of Humboldt-University Berlin
where the main part of the work was done.
V.G.B. is supported by RFBR grant 11-02-01227-a and by grant of the 
Russian Ministry of Science and Education.

\section*{Appendix: $SU(3)$ calorons}

The $SU(N)$ instantons at finite temperature (or calorons)
with non-trivial holonomy~\cite{Kraan:1998pm,
Kraan:1998sn,Lee:1998bb} can be considered as composites of $N$
constituent monopoles, seen only when the Polyakov loop at spatial
infinity (holonomy) is non-trivial.
In the periodic gauge,
$A_\mu(t\!+\!\cT,\vec x)\!=\!\!A_\mu(t,\vec x)$ it is defined as
\beq
\pl=\lim_{|\vec x|\rightarrow\infty}
P\,\exp(\int_0^\cT A_0(\vec x,t)dt).
\eeq
After a suitable constant gauge transformation, the Polyakov
loop can be characterised by real numbers $\mu_{m=1,...,n}$
($\sum_{m=1}^n\mu_m\!=\!0$) that describe the eigenvalues of the holonomy
\beqa
&&\plo=\exp[2\pi i\,{\rm diag}(\mu_1,\ldots,\mu_n)],\\
&&\mu_1\leq\ldots\leq\mu_n\leq\mu_{n+1}\!\equiv\!1\!+\!\mu_1.\nonumber
\eeqa
In units, where the inverse temperature $\cT=1$,
a simple formula for the $SU(N)$ action
density can be written~\cite{Kraan:1998pm,Kraan:1998sn} :
\beqa
&&\Tr F_{\mu\nu}^{\,2}(x)=\partial_\mu^2\partial_\nu^2\log\psi(x),\\
&&\psi(x)=\half\tr(\cA_n\cdots \cA_1)-\cos(2\pi t),\nonumber\\
&&\cA_m\equiv\frac{1}{r_m}\left(\!\!\!\bea{cc}r_m\!\!&|\vec y_m\!\!-\!
\vec y_{m+1}|\\0\!\!&r_{m+1}\eea\!\!\!\right)\left(\!\!\!
\bea{cc}c_m\!\!&s_m\\s_m\!\!&c_m\eea\!\!\!\right),\nonumber
\eeqa
with $r_m\!=\!|\vec x\!-\!\vec y_m|$ and $\vec y_m$ being the center of
mass radii of $m$ constituent monopoles, which can be assigned a mass
$8\pi^2\nu_m$, where $\nu_m\!\equiv\!\mu_{m+1}\!-\!\mu_m$. Furthermore,
$c_m\!\equiv\! \cosh(2\pi\nu_m r_m)$, $s_m\!\equiv\!\sinh(2\pi\nu_m r_m)$,
$r_{n+1}\!\equiv\! r_1$ and $\vec y_{n+1}\!\equiv\!\vec y_1$.

For $SU(3)$ calorons we correspondingly pa\-ra\-me\-tri\-ze the asymptotic
holonomy as
$\plo =
\mathrm{diag}(\mathrm{e}^{2\pi i\mu_1},\mathrm{e}^{2\pi i\mu_2},
\mathrm{e}^{2\pi i\mu_3})$,
with $\mu_1 \leq \mu_2 \leq \mu_3 \leq \mu_4 = 1+\mu_1$
and $\mu_1+\mu_2+\mu_3 = 0$.  Let $\vec{y}_1$, $\vec{y}_2$ and $\vec{y}_3$ be
three $3D$ position vectors of dyons remote from each other. Then
a caloron consists of three lumps carrying the instanton action split
into fractions $m_1 = \mu_2 - \mu_1$, $m_2 = \mu_3 - \mu_2$ and $m_3 = \mu_4 - \mu_3$,
concentrated near the $\vec{y}_i$.

Provided the constituents are well separated, the Polyakov loop values
at their positions $~\vec{y}_m,~m=1,2,3~$ are ~\cite{vanBaal:1999bz}
\beqa
\nonumber
\Pol(\vec{y}_1)&=
&\diag(\hphantom{-}e^{-\pi{i}\mu_3},\hphantom{-}e^{-\pi{i}\mu_3},\hphantom{-}
e^{2\pi{i}\mu_3}),\\  \label{eqn:polinterplay}
\Pol(\vec{y}_2)&=
&\diag(\hphantom{-}e^{2\pi{i}\mu_1},\hphantom{-}e^{-\pi{i}\mu_1},\hphantom{-}
e^{-\pi{i}\mu_1}),\\
\nonumber
\Pol(\vec{y}_3)&=
&\diag(-e^{-\pi{i}\mu_2},\hphantom{-}e^{2\pi{i}\mu_2},-e^{-\pi{i}\mu_2}).
\eeqa

\begin{figure}
\vspace{3 cm}
\includegraphics[width=.2\textwidth]{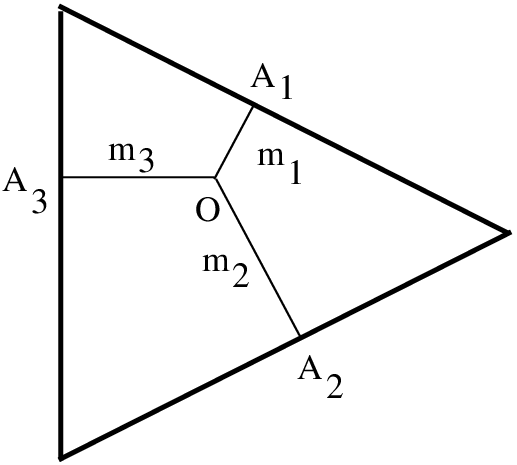}
\caption{ The regular triangle with inner point $O$  and three perpendiculars
to triangle sides. The sum of them $OA_1+OA_2+OA_3\equiv m_1 + m_2 + m_3 = 1$
is constant for all inner points.
}
\label{fig:triangle}
\end{figure}

The  complex numbers representing the trace of Polyakov loop
$PL = \frac{1}{3}\rm {Tr}\Pol$
occupy some region on the complex plane (see e.g. Fig. \ref{fig:purepolscat}).
The holonomy $\plo$ is traced to point $PL_{\infty}$ that is close
to (one third of) the trace of Polyakov loop averaged over all lattice points
and is near zero in the confining phase and near one in the deconfining phase.
There is a one-to-one correspondence between (one third of) the trace of
Polyakov loop and three numbers $m_1, m_2, m_3, m_1 + m_2 + m_3 = 1$ that
define the eigenvalues of $SU(3)$ matrix and hence its trace. Three numbers
can be represented by the inner point of regular triangle for which the sum
of the lengths of three perpendiculars to triangle sides is constant
(equal to one, see Fig. \ref{fig:triangle}). The region on the complex plane
occupied by (one third of) the trace of Polyakov loop can be considered as
some nonlinear deformation of this regular triangle. The point $O$ on
Fig. \ref{fig:triangle} corresponds to $PL_{\infty}$ while points
$A_1, A_2, A_3$ correspond to the values of
(one third of) the trace of the Polyakov loop at the constituent positions
(\ref{eqn:polinterplay}).

\bibliographystyle{apsrev}

\end{document}